\documentclass[prb,aps,reprint]{revtex4-1}
\usepackage{graphicx}
\usepackage{bm}
\begin{document}

\title{
Edge States and Stacking Effects in Nanographene Systems
}

\author{Kikuo Harigaya}
\email{k.harigaya@aist.go.jp}
\homepage{http://staff.aist.go.jp/k.harigaya/}
\author{Hiroshi Imamura}
\affiliation{Nanosystem Research Institute, AIST, Tsukuba 305-8568, Japan}

\author{Katsunori Wakabayashi}
\affiliation{National Institute for Materials Science,
Tsukuba 305-0044, Japan}

\author{Osman Ozsoy}
\affiliation{Erciyes University, Kayseri, Turkey}

\pacs{}

\begin{abstract}
Bilayer graphene nanoribbon with zigzag edge is investigated with the
tight binding model. Two stacking structures, $\alpha$ and $\beta$,
are considered.  The band splitting is seen in the $\alpha$ structure,
while the splitting in the wave number direction is found
in the $\beta$ structure.  The local density of states in the
$\beta$ structure tend to avoid sites where inter-layer hopping 
interactions are present.
\end{abstract}

\maketitle

\section{Introduction}

The graphite, multi-layer, and single-layer graphene materials 
have been studied intensively, since the electric field effect 
has been found in atomically thin graphene films.\cite{novo} These 
materials can be regarded as bulk systems. On the other hand, 
nanographenes with controlled edge structures have been predicted 
to have localized states along the zigzag edges.\cite{fujita}  
The presence of the edge states has been observed by experiments 
of scanning tunneling spectroscopy.\cite{kobaya,niimi}  Thus, the 
studies of the edge states are one of the interesting topic of 
the field.  The recent atomic bottom-up fabrication of nanoribbons
really promotes experimental and theoretical investigations.\cite{cai}

Previously, one of the present authors has studied the
stacking effects of the nanographene by considering weak
inter-layer hopping interactions in the tight binding model.\cite{hrgy1,hrgy2}
The cluster calculations have been performed, and compared
with experiments of the magnetic properties.  It has been
found that open shell nature of the electronic orbital
of the each layer is important.  The existence of the
spin in the each layer gives rise to the magnetism
of the stacked nanographene. Therefore, the simple
tight binding model is effective.

In this paper, bilayer graphene nanoribbon with zigzag edge 
is investigated with the tight binding model. Two stacking 
structures, $\alpha$ and $\beta$, are considered.  The band 
splitting is seen in the $\alpha$ structure, while the splitting 
in the wave number direction is found in the $\beta$ structure.  
The local density of states in the $\beta$ structure tend to 
avoid sites where inter-layer hopping interactions are present.

This paper is organized as follows.  In the next section,
the model is explained.  The results are shown in Section 3.
The paper is closed with summary in Section 4.

\begin{figure}[t]
\centerline{
\includegraphics[width=\columnwidth]{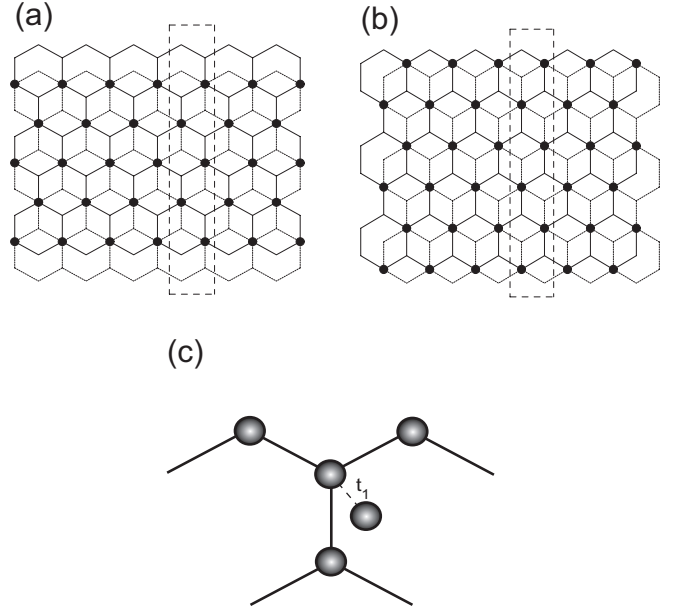}
}
\caption{
A-B stacked bilayer graphene nanoribbon with zigzag edges.
The upper layer is shown by the solid lines, and the lower layer 
by dotted lines.  In the $\alpha$ structure (a), the upper layer 
is shift by the bond length downward to the position of the 
lower layer.  The region surrounded by the dashed line is
the unit cell in the direction of the one dimensional
direction.  At the circles, two carbon atoms of the upper
and lower layers overlap completely, and there is the
weak hopping interaction $t_1$ here.  In (b), the $\beta$
structure is shown, where the lower layer is shift
right-down direction so the stacking pattern is different
from that of the $\alpha$ structure. The convention of
the $\alpha$ and $\beta$ structures are the same used in the
paper.\cite{lima} (c) The detailed
view around the edge atom in the $\beta$ structure is
displayed.
}
\label{fig:Fig1}
\end{figure}

\section{Model}

We consider the following tight binding model
\begin{eqnarray}
H=-t \sum_{\langle i,j \rangle, \sigma} 
(c_{i,\sigma}^\dagger c_{j,\sigma} + {\rm H.c.}) \nonumber \\
-t \sum_{\langle i,j \rangle, \sigma}
(d_{i,\sigma}^\dagger d_{j,\sigma} + {\rm H.c.}) \nonumber \\
-t_1 \sum_{\langle i,j \rangle, \sigma} 
(c_{i,\sigma}^\dagger d_{j,\sigma} + {\rm H.c.}),
\end{eqnarray}
where $c_{i,\sigma}$ and $d_{i,\sigma}$ are the annihilation
operators of electrons at the lattice site $i$ of the spin $\sigma$
on the upper and lower layers, respectively.  The quantity $t$
is the hopping integral of $\pi$ electrons between neighboring
lattice sites.  Two stacking patterns,
shown in Figs. 1 (a) and (b), are considered.  They are named
as $\alpha$ and $\beta$ structures, respectively.  This 
convention has been used in the literature.\cite{lima}
There are $N_z=6$ zigzag lines in upper and lower layers.
The detailed view near the edge atom in the $\beta$
structure is displayed in Fig. 1 (c).  There is a weak hopping
integral $t_1$ along the dashed line.

%========================================
% Band
%========================================
\begin{figure}[t]
\centerline{
\includegraphics[width=\columnwidth]{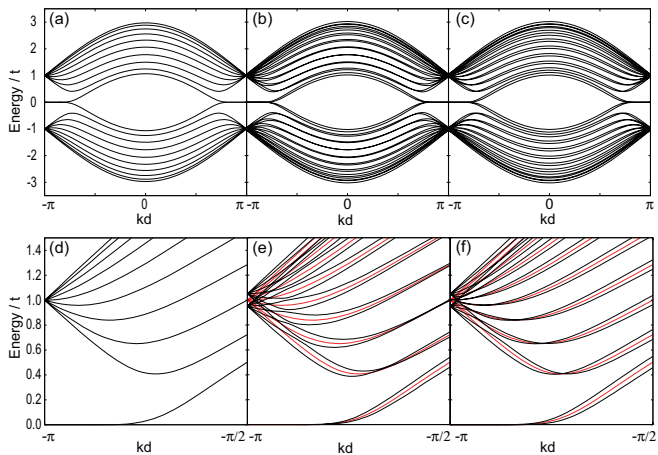}
}
\caption{
Energy band structures for the single layer (a), 
$\alpha$ structure (b), and $\beta$ structure (c).
Details near the Brillouin zone edge are magnified
 for the single layer (d), 
$\alpha$ structure (e), and $\beta$ structure (f).
In (e) and (f), the energy bands of the single
layer are shown by the red lines for comparison.
}
\label{fig:Fig2}
\end{figure}

\begin{figure}[t]
\centerline{
\includegraphics[width=\columnwidth]{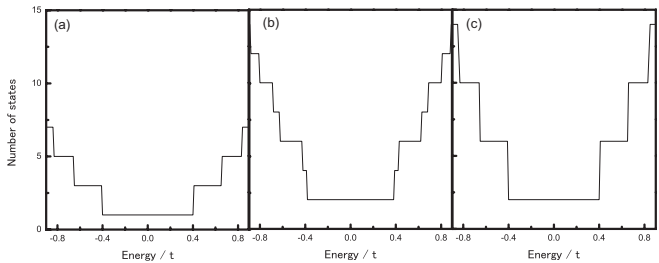}
}
\caption{
Number of states against the energy for the
single layer (a), $\alpha$ structure (b),
and $\beta$ structure (c).
}
\label{fig:Fig3}
\end{figure}

\section{Results}

In order to see effects of inter-layer interactions,
the tight binding model is solved numerically.
The model is solved using the Bloch theory
of the band calculation.
Energy band structures of the $N_z=20$ systems
are displayed in Fig. 2, for the single layer (a), 
$\alpha$ structure (b), and $\beta$ structure (c).
The interlayer interaction strength is $t_1=0.1t$.
In (b), the split of the energy bands is seen
compared with Fig. (a).  In contrast, split in the
perpendicular direction is small in Fig. (c).
In order to see split structures clearly,
details near the Brillouin zone edge are magnified
 for the single layer (d), $\alpha$ structure (e), 
and $\beta$ structure (f).  In (e) and (f), the 
energy bands of the single layer are shown by the 
red lines for comparison.  The nearly flat band
due to the edge state in $-\pi < kd < -2\pi/3$
is present at the energy $E \sim 0$ in (d),
where $d$ is the unit cell length of the one
dimensional direction of Figs. 1 (a) and (b).
The energy bands starts at $E=1.0t$ at $kd=\pi$,
typical to the graphene structure.  In (e),
the energy split is magnified again.  However,
in (f), the energy split is not seen,
and split in the wave number direction is found.
This property is confirmed by looking at the numerical
data, also.

It is interesting to look at how such the difference of
the band split in the $\alpha$ and $\beta$ structures
appear, in the quantization of conductance.
Number of states in the energy window for the
positive wave number is calculated
for the single layer, $\alpha$, and $\beta$ structures.
Fig. 3 displays the calculated results for the
parameters $N_z=20$ and $t_1=0.1t$ again.
Around the energy $E=0$ of the single layer (a),
the energy band is singly degenerate, as shown
in Fig. 2 (d).  At the energy $E=0.4t$, the energy
window reaches the first parabolic band, and the
number of states jumps by two.  The energy window
reaches the second parabolic band at the energy
$E=0.65t$.  In this way, a series of jumps by two
is realized in (a).  For the $\alpha$ structure (b),
number of states near the energy $E=0$ is two
due to the bilayer property.  There are two
steps of the jump of two around the energy $E=0.4t$,
owing to the energy split of the first parabola.
Also, two steps are present closely around the
energy $E=0.7t$ due to the second parabolic band.
On the other hand, each jump of the number
of states becomes four for the $\beta$ structure (c)
due to the split in the wave number direction.
The dependence of the energy reflects the band 
structures, and this will appear quantization
of conductance experimentally.

\section{Summary}

In summary, weak inter-layer interactions have been considered
for the bilayer graphene nanoribbon with zigzag edge.
The $\alpha$ and $\beta$ stacking structures have been considered.  
The band splitting is seen in the $\alpha$ structure, while the 
splitting in the wave number direction is found in the $\beta$ 
structure.  The local density of states in the $\beta$ structure 
tend to avoid sites where inter-layer hopping interactions exist.

\end{document}